\documentstyle[aps,epsfig,epsf,psfig]{revtex}
\def\be{\begin{equation}}
\def\ee{\end{equation}}
\def\bearr{\begin{eqnarray}}
\def\eearr{\end{eqnarray}}
\def\zbf#1{{\bf {#1}}}
\def\bfm#1{\mbox{\boldmath $#1$}}

\begin{document}
\draft
\preprint{}

\title{Vacuum structure in QCD and Correlation functions\footnote{
Talk prepared for DAE Nuclear Physics Symposium, December 27th--31st,
 1999,Chandigarh, India}}
\author{
Hiranmaya Mishra \footnote {email address hm@prl.ernet.in}}
\address{Theory Division, 
Physical Research Laboratory, Navrangpura, Ahmedabad 380 009, India}
\maketitle
\begin{abstract}
We discuss here a model
of QCD vacuum in terms of quark antiquark and gluon condensates
alongwith their fluctuations.
The correlation functions of hadronic currents in such a vacuum are
evaluated to extract hadron properties. The presence of fluctuations 
of the condensates are emphasized. The structure of vacuum is then 
generalised to finite temperatures to study correlation functions at 
finite temperatures.  Finally, we discuss the vacuum structure at 
finite densities with diquark condensates in a Nambu JonaLasinio type model. 
The coupled mass gap and superconducting gap equations are solved 
selfconsistently. For certain parameters of the model
nontrivial solutions of both the gap equations are obtained. The 
equation of state is also computed.
\end{abstract}
\vskip 0.5cm
 \section{Introduction}
Quantum chromodynamics (QCD) is now accepted to be the theory of strong 
interaction in terms of quarks and gluons, and, at a secondary level,
of hadrons. It is a nonabelian gauge theory with $SU(3)$ as the gauge group. 
This nonabelianess of the interaction leads to two important consequences. 
Firstly, the interaction become weak at high momentum (above several GeV) 
transfer processes where perturbation theory is applicable. At low energies 
and momenta, relevant to most of nuclear physics, the quark gluon coupling 
strength becomes large and an expansion in powers of this coupling is not 
useful. The basic difficulty appears
to be an understanding of the ground state properties of QCD or its vacuum
structure which plays an important role for the related physics \cite{feyn}.

Conceptually, low energy QCD has many common feature with condensed matter
physics. The vacuum here appear to be consisting of having fluctuating quarks
and gluon fields with average properties being described by condensates of
quarks and gluons. The quark condensate $\langle \bar q q\rangle$, i.e. expectation value of quark scalar densities, plays an important role in the context of strong interactions. It displays the breaking of chiral symmetry. This symmetry
is crucial to our understanding of nucleons,nuclei and dense matter. The other quantity that charcterise the QCD vacuum is the gluon condensate. The energy density of vacuum is lowered by presence of electric and magnetic gluon fields
in the ground state. These condensates were introduced in the context of QCD sumrules and the values are estimated from the charmonia spectroscopy as
$\langle \alpha_s/\pi GG\rangle=\langle 2\alpha_s/\pi^2
(\zbf E^2-\zbf B^2 )\rangle= 0.012 GeV^4$ and $\langle\bar q q\rangle=
-(250 MeV)^3$ \cite{svz}.

Thus QCD vacuum poses a rich complicated many body problem. As in 
other many body system ground state correlation functions give an 
insight to the ground state structure. One might be interested in
 asking questions like what happens to a meson pair when placed in
 such a condensate medium or what happens to
interquark interaction as a function of their spatial separation.
We might remind ourselves that nucleon scattering phase shifts gives 
information regading inter-nucleon forces complementary to that infered 
from their bound state namely from the properties of deuteron. N-N 
scattering allows one to study different components of nuclear
 forces (spin-spin, tensor,$\cdots$) at different spatial separation in
 much more detail than deuteron observables
which is an composite effects of all channels. In QCD we however do not have 
free quarks or gluons due to confinement. None the less, one can infer about
inter-quark interaction as a function of their separation  in different
channels by studying propagators and correlation functions of hadron currents
in  a more detailed manner than the composite effects reflected in hadron
bound state.
 The correlation functions that we shall consider here are spacelike
separated correlation functions. To be precise we shall take the
correlation functions to be defined at equal time so that their separation is 
purely spatial. Such correlation functions have several appealing features.
They describe different physics at different spatial separations; they
can be calculated in some channels phenomenologically from $e^+e^-
\rightarrow $hadrons data or from $\tau$ decay experimental data. Moreover they can be evaluated in lattice simulations \cite{neglecor} or in some 
model for vacuum\cite{surcor}.

Our appraoch here shall be assuming a structure for the vacuum and then
examining its consequences regarding correlator phenomenology. More precisely,
we shall use phenomenological results of correlation functions of hadronic currents \cite{neglecor} to guide us towards a ``true" structure of QCD vacuum.
We organise this note as follows. In section {\bf II}, we shall discuss a
construct for the vacuum in terms of quark and gluon condensates \cite{prliop},
and the resulting correlation functions \cite{propcor}. It appears that
condensates alone do not give rise to correct phenomenology of correlators
unless one includes fluctuations of such condensates \cite{plb}. Inclusion of
the fluctuation fields yields correlation functions consistent with 
the phenomenology of correlation functions. Such a structure of vacuum
of zero temeperature is then generalised to finite temperature in section 
{\bf III}. Here we shall also compute the finite temperature correlators
to determine temperature dependent hadron properties \cite{cort}.
In section {\bf IV},
we shall discuss the vacuum structure at high density and the QCD vacuum with
diquark condensate giving rise to color superconductivity.
 Finally, we summarize our results with some 
remarks and discussions in section {\bf V}.

\section{An ansatz for the QCD vacuum and correlation functions 
}

A variational ansatz was considered in Ref.\cite{prliop} for the QCD vacuum
with an explicit construct involving quark antiquark pairs {\it and} 
gluon pairs. The  trial ansatz was given as
\be
|vac\rangle=\exp(B_F^\dagger-B_F)(B_G^\dagger-B_G)|0\rangle
\label{vac}
\ee
where, the Bogoliubov pair creation operators for the quarks and the 
gluons are given respectively as
\begin{mathletters}
\be
B_F^\dagger=
\int q_I^0(\zbf k)^\dagger(\bfm {\sigma }\cdot  \zbf k)
h(\zbf k) \tilde q_I^{0} (\zbf k)d\zbf k,
\label{bf}
\end{equation}
and,
\be
B_G^\dagger=
\int a_i^a(\zbf k)^\dagger g(\zbf k) a_i^{a} (-\zbf k)d\zbf k.
\label{bg}
\end{equation}
\end{mathletters}
In the above $q^0_I,\tilde q^{\dagger 0}_I$ are two component quark 
 and antiquark annihilation operators respectively. The subscript $0$
indicates that they annihilate the peruturbative vacuum $|0\rangle$
i.e. $q_I^0|0\rangle=0=\tilde q^\dagger|0\rangle$.
$a_i^a$ is the gluon annihilation operator. The operators satisfy the
quantum algebra given in Coulomb gauge as
\begin{mathletters}
\be
\left [q^{i0}_{Ir}(\zbf k),q^{0j}_{Is}(\zbf k')^\dagger\right]_+=
\delta^{ij}\delta_{rs}
\delta(\zbf k-\zbf k'),
\ee
and,
\be
\left [a_i^a(\zbf k),a_j^b(\zbf k')\right ]=\delta^{ab}\left(\delta_{ij}-
\zbf k_i\zbf k_j/\zbf k^2\right)\delta(\zbf k-\zbf k').
\ee
\end{mathletters}

\noindent Finally, $h(\zbf k)$ and $g(\zbf k)$ are two trial
 functions associated with
the quark antiquark condensates and gluon pairs respectively.
Clearly, a construct as in Eq.(\ref{vac})has  an obvious parallel to
BCS theory of superconductivity.
Such a structure for vacuum eventually reduces to Bogoliubov transformation
for the operators. Then one can calculate the, energy functional--the
 expectation value of the Hamiltonian  which is a functional of the 
condensate functions.  Since the functions cannot be determined through 
functional minimisations one can parametrise the condensate functions with
some trial functions simple enough to manipulate numerically as well as 
reasonable enough to simulate correct physical behaviour. In Ref.\cite{prliop},
the following choices were made (with $ k=|\zbf k|$)
\begin{mathletters}
\be
\tan 2h(\zbf k)=\frac{A'}{(\exp(R^2k^2)-1)^{1/2}}
\ee
\be
\sinh g(\zbf k)=A\exp(-bk^2)
\ee
\end{mathletters}

In Ref.\cite{prliop} the energy density was minimised with respect to the condensate parameters subjected to the constraint that the pion decay constant
$f_\pi$ and the gluon condensate value $(\alpha_s/ \pi)\langle GG\rangle$
turns out to be the experimental values of 93 MeV and $0.012GeV^4$ respectively.
The result of such a minimisation then leads to instabilty 
of perturbative vacuum to condensate formation beyond a 
critical coupling of $\alpha_s^c=0.6$.
For $\alpha_s=1.28$, the charge radius of pion comes out correctly.
The values of $A'$ and $R$ turns out to be $A'\simeq 1 $ and $R\simeq 0.96fm.$
Further some of the baryonic properties like charge radius of proton, magnetic
moments of proton and neutron turns out to be close to their corresponding
 experimental values. Further, the bag constant-- the energy 
difference between the perturbative and the nonperturbative vacuum
 turns out to be $\epsilon_0=-(140 MeV)^4$which appears to be in 
 general agreement with the phenomenological value of this parameter.

With such a description of the vacuum in terms of condensates let us look
at the correlation functions and propagators in a condensed medium. The
equal time propagator is given as\cite{mac}
\begin{eqnarray}
S_{\alpha\beta}(\zbf x)& = & \langle \frac{1}{2}\left[q_\alpha(\zbf x),
\bar q_\beta(\zbf 0)\right ]\rangle\nonumber\\
&=& \frac{1}{2}\frac{1}{(2\pi)^3} \int e^{i\zbf k\cdot\zbf x}
\left[ \sin 2h(\zbf k)-\bfm\gamma\cdot \zbf k \cos 2h(\zbf k)\right ]\\
\label{prop}
\end{eqnarray}
Clearly, free massless propagator corresponds to $h(\zbf k)\rightarrow 0$
limit of above equation and is given as
 $S_0=-(i/2\pi^2)(\bfm \gamma\cdot \zbf x/x^4)$. 
Different components of the propagator
given in Eq.(\ref{prop}) can be analysed\cite{propcor}
 and it turns out to be qualitatively similar to those obtained from
other nonperturbative calculations like an instanton liquid model for
 QCD vacuum\cite{surcor,suprop}. Further, a small x expansion of Eq.(\ref{prop})
yields the propagtor as that one would get in the operator product expansion in the vacuum saturation approximation\cite{suprop}.

We shall next consider the correlation functions of mesonic currents of generic
form $J(\zbf x)=\bar q_\alpha(\zbf x)\Gamma_{\alpha\beta}q_\beta(\zbf x)$,
where $\alpha,\beta$ are spinor indices and $\Gamma$ is a 4$\times$4 matrix
in Dirac space $(1,\gamma_5,\gamma_\mu,\gamma_\mu\gamma_5)$. The
 equal time correlation function is defined as
\be
R(\zbf x)=\frac{1}{2}\langle vac|J(\zbf x)\bar J(\zbf 0)|vac\rangle
+\langle vac|\bar J(\zbf 0) J(\zbf x)|vac\rangle
\label{rx}
\ee
With the definition of condensate vacuum as in Eq.(\ref{vac}) and
the propagator given in Eq.(\ref{prop}), Eq.(\ref{rx}) reduces to
\be
R(x)=-Tr[S(x)\Gamma'S(-x)\Gamma]
\label{rx1}
\ee
where, $\Gamma'=\gamma_0\Gamma^\dagger\gamma_0$.
We shall be normalising the correlation functions with that of free
massless correlation function which is given as parallel to Eq(\ref{rx1})
as
\be
R_0(x)=-Tr[S_0(x)\Gamma'S_0(-x)\Gamma]
\ee 
The ratio $R(x)/R_0(x)$ can then be evaluated for different channels.
It turns out that  correlation function so obtained has similar
qualitative behaviour as one can obtain from phenomenology in all
chnnels except for the pseudoscalar channel \cite{propcor}. 
Phenomenologically, in this channel there is a strong attraction with 
the ratio becoming about 100 at a separation of about a fermi\cite{surcor}.
The calculated
value however turns out to be as low as 1.2 around that value. All these
results depend very weakly on the functional form of $h(\zbf k)$.

In view of this outcome, it is obvious that some crucial physics is missing
from the model of vacuum considered in Eq.(\ref{vac}) and has to be
 supplemented by additional effects. In the present framework this means that
quark propagators
alone do not describe the correlation functions and there has to be
contributions from irreducible four point structure of the vacuum. This can be 
thought of as a manifestation of fluctuation of the condensates. Thus we have

\begin{eqnarray}
T\bar q_\alpha(\zbf x)q_\beta(\zbf x)q_\gamma(\zbf 0)q_\delta(\zbf 0)& =&
S_{\beta\gamma}(\zbf x)S_{\delta\alpha}(-\zbf x)+:
\bar q_\alpha(\zbf x)q_\beta(\zbf x)q_\gamma(\zbf 0)q_\delta(\zbf 0):
\nonumber\\
&=&
S_{\beta\gamma}(\zbf x)S_{\delta\alpha}(-\zbf x)+\Sigma_{\beta\gamma}(\zbf x)
\Sigma_{\delta\alpha}(-zbf x)\\
\label{flu}
\end{eqnarray}
where, we have introduced the composite fields $\Sigma$ to include the
 effects of fluctuation with $\langle vac|\Sigma\Sigma|vac\rangle=0$ but
$\langle\Omega|\Sigma\Sigma|\Omega\rangle\ne 0$, where, $|\Omega\rangle$is
the ``new improved" QCD vacuum including the condensate fluctuations.
The correlation function then takes the form

\be
R(x)=-\left[Tr[S(x)\Gamma'S(-x)\Gamma]
+Tr\left[\Sigma(\zbf x))\Gamma'\Sigma(-x)\Gamma\right]\right]
\label{fcor}
\ee
The structure of $\Sigma$ field should be such that it contributes mostly to
the pseudoscalar channel and should not affect the other channels very much.
Such a condition restricts the composite field to be of the form
\be
\Sigma_{\alpha\beta}(\zbf x)=\mu_1^2(\gamma^i\gamma^j\epsilon_{ijk}\phi^k
(\zbf x)+
\mu_2^2\delta_{\alpha\beta}\phi(\zbf x)
\label{sig}
\ee
where, we have introduced the scalar and vector fileds $\phi$ and $\phi^k$
such that 
\begin{mathletters}
\be
\langle\Omega|\phi^i(\zbf x)\phi^j(\zbf 0)|\Omega\rangle=\delta^{ij}g_V(\zbf x)
\ee

\be
\langle\Omega|\phi(\zbf x)\phi(\zbf 0)|\Omega\rangle=g_S(\zbf x)
\ee
\end{mathletters}

From general considerations, we may write down the functions $g_V(\zbf x),
g_S(\zbf x)$ as\cite{plb}
\begin{mathletters}
\be
g_V(\zbf x)=\frac{1}{2\pi^2 x}
\left [\mu_1K_1(\mu_1 x)-\mu_3K_1(\mu_3 x)\right]
\label{vflu}
\ee
\be
g_S(\zbf x)=\frac{1}{2\pi^2 x}\left [\mu_2K_1(\mu_4 x)-\mu_5K_1(\mu_6 x)\right]
\label{sflu}
\ee
\end{mathletters}

Using Eq.s(\ref{fcor},\ref{sig},\ref{vflu},\ref{sflu}) one can 
have the expression for the
correlation functions in different channels. 
Explicitly different currents in different channels and
the contributions from the propagator and the fluctuation
fields are shown separately in Table {\ref{table1}.
We have also included here the currents for the baryons.
The resulting correlation functions normalised to correlation functions 
obtained by treating the quarks as massless and noninteracting 
are plotted in Fig.\ref{corfig}.
\begin{figure}[h]
\epsfysize=16cm
\epsfbox[20 40 490 630]{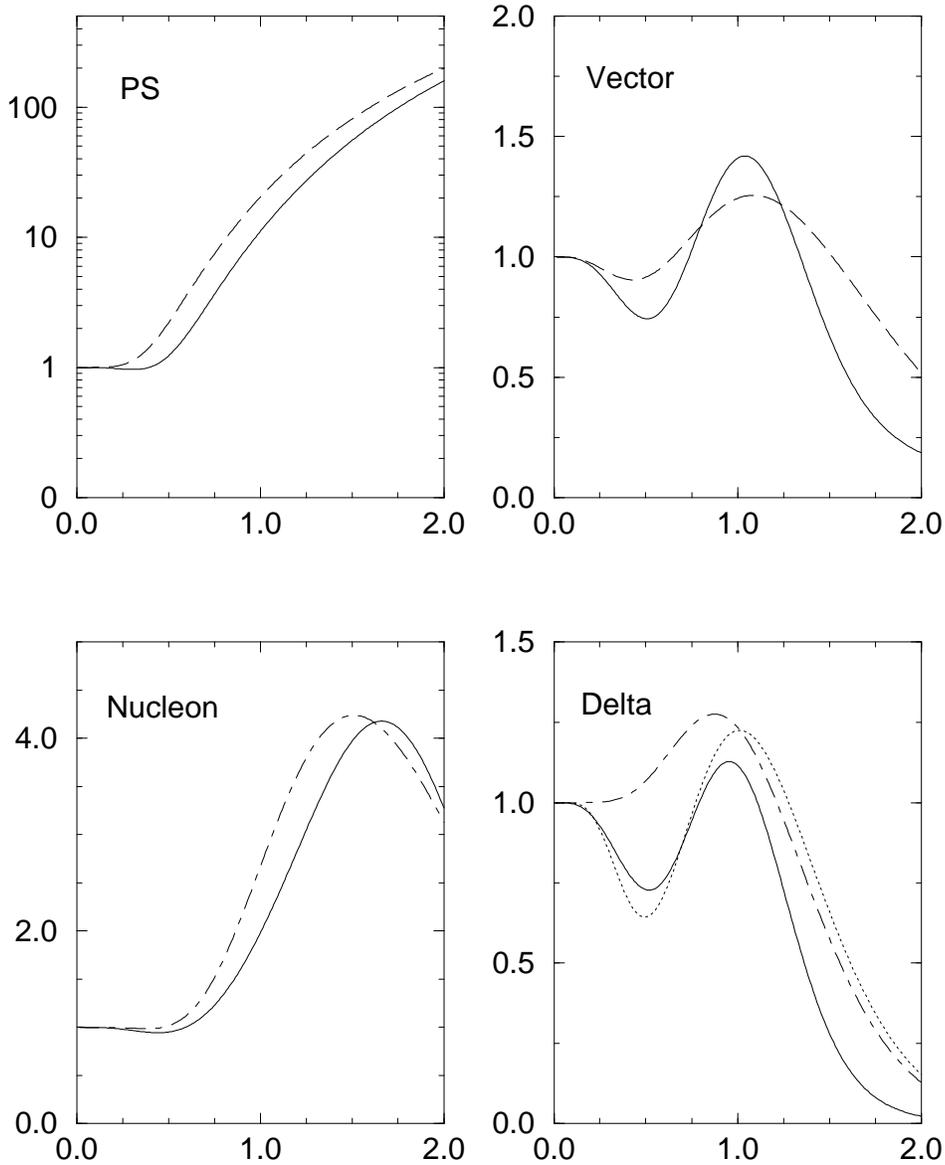}
\caption{\em The ratio of hadron correlation functions in QCD
vacuum to the correlation functions for noninteracting massless quarks
$R(x)/R_0(x)$ , vs distance x (in fermi). Our results are given by
the solid curves. The emperical results determined from dispersion 
analysis of experimantal data in Ref. are shown by 
long dashed lines. The results from lattice calculations and 
instanton liquid model are denoted by dotted and dot-dashed lines
respectively.}
\label{corfig}
\end{figure}

\begin{table}
\caption{Meson currents and correlation functions \label{table1}}
\begin{tabular}{llll}
CHANNEL & CURRENT & \multicolumn{2}{c}{CORRELATION FUNCTIONS
 $\displaystyle \left[ \; \frac{R(x)}{R_{0}(x)} \; \right]$ }\\
 & & Without fluctuations\tablenotemark[1]&Fluctuation contribution\\   
 & & &(Vector($ F^V$) and Scalar ($ F^S$)) \\
 & & & \\ 
\tableline
Vector & $\bar u \gamma_{\mu} d$  
& $[F(x)]^2 + \frac{\pi}{4} \frac{x^6}{R^6} e^{-x^2/R^2}$ 
&$ F^V= 0$  \\
 & & & $ F^S= 8 \pi^4 x^6 g_S(2x)$ \\ 
\hline  
Pseudoscalar & $\bar u \gamma_{5} d$  
&  $[F(x)]^2 + \frac{\pi}{8} \frac{x^6}{R^6} e^{-x^2/R^2}$
&$ F^V= -48\pi^4 x^6 g_V(2x)$  \\
 & & & $ F^S= 4 \pi^4 x^6 g_S(2x)$ \\ 
\hline  
Nucleon  & $ \epsilon_{abc} \left[\tilde u^a(x) C \gamma_\mu u^b(x)
\right] \gamma^\mu \gamma_5 d^c(x) $
& $ [F(x)]^3 + \frac{\pi}{16} \frac{x^6}{R^6}  e^{-x^2/R^2} F(x) $
 &$ F^V= -4\pi^4 x^6 g_V(2x) F(x)$  \\
 & & & $ F^S= 2 \pi^4 x^6 g_S(2x) F(x)$ \\ 
\hline  
Delta & $
\epsilon_{abc} \left[\tilde u^a(x) C 
\gamma_\mu u^b(x) \right]u^c(x) $
& $ [F(x)]^3 + \frac{\pi}{4} \frac{x^6}{R^6}  e^{-x^2/R^2} F(x) $
&$ F^V= 0$  \\
 & & & $ F^S= 8 \pi^4 x^6 g_S(2x) F(x)$ \\ 
\end{tabular}
\tablenotetext[1]{ $F(x)=\left[1+\frac{1}{2} x^2 I(x) \right] $
where $I(x) = \int_{0}^{\infty } \left( \cos k x - \frac{\sin k
x}{kx} \right) \frac{k e^{-R^2 k^2}}{1+(1-e^{-R^2 k^2})^{1/2}} dk $}
\end{table}
 To get the hadron parameters from the correlation functions
one relates the correlation function through a dispersion 
relation to a spectral density function. For example in the vector 
channel the dispersion
relation for the correlation function reduces to \cite{surcor}
\be
\Pi_{\mu\mu}(x)\equiv \langle \Omega|T(J_\mu(x)\bar J_\mu(0))|\Omega\rangle
= \frac{1}{4\pi^2}\int_0^{\infty} ds R_i(s)D(s^{1/2},x)
\label{disp}
\ee
with, the function $D(m,x)=(m/4\pi^2 x)K_1(mx)$ is a propagator of a particle
of mass $m$. The function $R_i(x)$ is the normalised spectral density function
related to the cross section of $e^+e^-$ annihilation to hadrons. In presence
of interaction, in general we do not know how to calculate the spectral density function . However we do know their qualitative behaviour based on experimental
information. To extract hadron parameters one parametrises the spectral function and then determine the parameters. A useful parametrisation valid at
zero temperature is in terms of a  Dirac delta function at the pole
 mass of the hadron accompanied by a step function continuum
at higher energy. Specifically,
for $\rho$ meson the spectral density function is parametrised by
\be
R_\rho(s)=3 \lambda^2\delta(s-M^2)+\frac{3s}{4\pi^2}\theta(s-s_0).
\label{rhodis}
\ee
\begin{table}
\caption{Fitted Parameters \label{table2}}
\begin{tabular}{lllll}
CHANNEL & SOURCE & M (GeV) & $\lambda$ & $\sqrt{s_0}$(GeV) \\
\tableline
Vector & Ours & 0.78 $\pm$ 0.005 & (0.42 $\pm$ 0.041 GeV)$^2$
& 2.07 $\pm$ 0.02 \\
 & Lattice & 0.72 $\pm$ 0.06 & (0.41 $\pm$ 0.02 GeV)$^2$
& 1.62 $\pm$ 0.23 \\
 & Instanton & 0.95 $\pm$ 0.10 & (0.39 $\pm$ 0.02 GeV)$^2$
& 1.50 $\pm$ 0.10 \\
 & Phenomenology & 0.78 & (0.409 $\pm$ 0.005 GeV)$^2$
& 1.59 $\pm$ 0.02 \\
\hline  
Pseudoscalar & Ours & 0.137 $\pm$ 0.0001 & (0.475 $\pm$ 0.015 GeV)$^2$
& 2.12 $\pm$ 0.083 \\
 & Lattice & 0.156 $\pm$ 0.01 & (0.44 $\pm$ 0.01 GeV)$^2$
& $\; \; \;<$ 1.0  \\
 & Instanton & 0.142 $\pm$ 0.014 & (0.51 $\pm$ 0.02 GeV)$^2$
& 1.36 $\pm$ 0.10 \\
 & Phenomenology & 0.138 & (0.480 GeV)$^2$
& 1.30 $\pm$ 0.10 \\
\hline  
Nucleon & Ours & 0.87 $\pm$ 0.005 & (0.286 $\pm$ 0.041 GeV)$^3$
& 1.91 $\pm$ 0.02 \\
 & Lattice & 0.95 $\pm$ 0.05 & (0.293 $\pm$ 0.015 GeV)$^3$
& $\; \; \;<$ 1.4 \\
 & Instanton & 0.96 $\pm$ 0.03 & (0.317 $\pm$ 0.004 GeV)$^3$
& 1.92 $\pm$ 0.05 \\
& Sum rule & 1.02 $\pm$ 0.12 & (0.337 $\pm$ 0.0.014 GeV)$^3$
& 1.5\\
 & Phenomenology & 0.939 & $ \; \; \; ?$
& 1.44 $\pm$ 0.04 \\
\hline  
Delta & Ours & 1.52 $\pm$ 0.003 & (0.341 $\pm$ 0.041 GeV)$^3$
& 3.10 $\pm$ 0.008 \\
 & Lattice & 1.43 $\pm$ 0.08 & (0.326 $\pm$ 0.020 GeV)$^3$
& 3.21 $\pm$ 0.34 \\
 & Instanton & 1.44 $\pm$ 0.07 & (0.321 $\pm$ 0.016 GeV)$^3$
& 1.96 $\pm$ 0.10 \\
 & Sum rule & 1.37 $\pm$ 0.12 & (0.337 $\pm$ 0.014 GeV)$^3$
& 2.1  \\
 & Phenomenology & 1.232  & $ \: \; \;?$
& 1.96 $\pm$ 0.10 \\
\end{tabular}
\label{tabpar}
\end{table}
\noindent where, M is the bound state mass, $\lambda$ 
is the coupling of the current
to the bound state and $s_0$ is the threshold for  continuum contribution.
The fitted parametrs to the correlation functions obatined
by us (solid curve in Fig.\ref{corfig} are given in Table \ref{tabpar}

As may be evident from this the results are
comparable with that of lattice calculation \cite{neglecor} 
and instanton liquid model calculations \cite{surcor}. Thus to be 
consistent with the data QCD vacuum must not only 
have condensates but also have their fluctuations.

\section{QCD vacuum and correlation functions at finite temperature}

As is well known \cite{lman} the QCD vacuum state changes with temperature. 
Lattice Monte Carlo simulations suggest that chiral symmetry is restored
around 150 MeV.  Thus it is interesting to look at correlation functions 
at finite temperature in the
context of behaviour of hadrons around the chiral phase transition
\cite{hotcor,shuhotcor}. It may be 
noted that there is little phenomenological information in this regime 
but there are several theoretical studies \cite{boch,hatsuda,brown,hatsuda2} 
using operator product expansion (OPE) and sum rule methods as well as using 
instanton liquid model for QCD ground state \cite{surcor2,vel}. In particular,
we shall generalise the method considerd in the previous section to include
temperature effects. The zero temperature vacuum defined in Eq.(\ref{vac})
can be generalised to finite temperature using the methodology of 
thermofield dynamics.
Here the thermal average of an operator 
is obtained as an expectation value of
the operator over the thermal vacuum\cite{umezawa}.
The thermal vacuum is obtained
from the zero temperature vacuum by a thermal Bogoliubov
transformation in an extended Hilbert space involving extra field
operators (thermal doubling of operators)\cite{umezawa}. Explicitly,
the thermal vacuum is given as
\be
|vac,\beta\rangle=\exp\left(\int d\zbf{k}\theta({\zbf k,\beta})
({q_I}^\dagger(\zbf k)\underline {q_I}^\dagger(-\zbf k)+
{\tilde q}_I(\zbf k)\underline{\tilde q}_I(-\zbf k))
- h.c.\right )|vac\rangle
\label{tvac}
\ee
where, the underlined oprators are the operators corresponding to
extra of Hilbert space. Further, the $q_I's$ are the opeartors
refers to the fact that they are the quasi paricle operators 
corresponding to basis defined by the vacuum of Eq.(\ref{vac}) i.e.
$q_I|vac\rangle=0=\tilde q^\dagger_I|vac\rangle$ and finally 
$\theta(k,\beta)$ is the function for the thermal Bogoliubov 
transformation and is related to the number density function 
given as
\be
\sin^2\theta(\zbf k,\beta)=\frac{1}{\exp(\beta\epsilon)+1}
\ee
where, $\epsilon(\zbf k)$ is the single particle energy
given as $\epsilon(\zbf k)=\sqrt{\zbf k^2+m(k)^2}$. In the presence of 
condensate  the dynamical mass is given as
$m(k)=k \tan 2h(k)$ \cite{prliop}. As before the 
equal time propagator can be calculated including the
temperature effects as
\begin{eqnarray}
S_{\alpha\beta}(\zbf x)& = & \langle  vac,\beta|\frac{1}{2}
\left[q_\alpha(\zbf x),
\bar q_\beta(\zbf 0)\right ]|vac,\beta\rangle\nonumber\\
&=& \frac{1}{2}\frac{1}{(2\pi)^3} \int e^{i\zbf k\cdot\zbf x}
\cos\!2\theta(\zbf k,\beta)\left[
\sin 2h(\zbf k)-\zbf\gamma\cdot \zbf k \cos 2h(\zbf k)\right ]\\
\label{propt}
\end{eqnarray}

As in Section III, we shall take a gaussian ansatz for
the condensate function $\sin 2h(\zbf k)=\exp(-R(T)^2\zbf k^2/2$,
with the condensate scale parameter $R$, now being temperature dependent.
In order to determine $R(T)$ or equivalently the ratio
$S(T)=R(T=0)/R(T)$, we
first evaluate our expression of the order parameter 
(the condensate value) at finite temperature. In terms of the
dimensionless variable $\eta=Rk$, this is given as
\be
\frac{\langle \bar q q\rangle_T}{\langle \bar q q\rangle_{T=0}}
= S(T)^3\bigg[ 1-2\sqrt{\frac{2}{\pi}}\int
e^{-\eta^2/2} \sin^2(z,\eta) \eta^2 d\eta \bigg],
\label{qqrat}
\ee
where $\sin^2\theta(z,\eta)={1}/\left (\exp(z\epsilon (\eta))+1)\right)$, with 
$z=\beta/R(T)$ and $\epsilon(\eta)=\eta/\cos 2h(\eta)$.

We can obtain  $S(T)=R(T=0)/R(T)$ if we know the temperature
dependence of the order parameter
on the left hand side of Eq.(\ref{qqrat}).
As there are no phenomenological inputs for this, we shall 
proceed in the following manner to take the temperature
dependence of the quark condensate. For low temperatures
we shall take the results from chiral perturbation theory 
(CHPT) which is expected to be valid at least for
small temperatures\cite{chpt}. For higher temperatures near the critical 
temperature, lattice simulations seem to yield the universal 
behaviour\cite{lman} with a large correlation length associated with 
a second order phase transition
for two flavor massless QCD. We shall use such a critical behaviour to
consider the temperature dependence of the order parameter near the 
critical temperature. 
\begin{figure}
\epsfysize=10cm
\epsfbox[50 275 540 650]{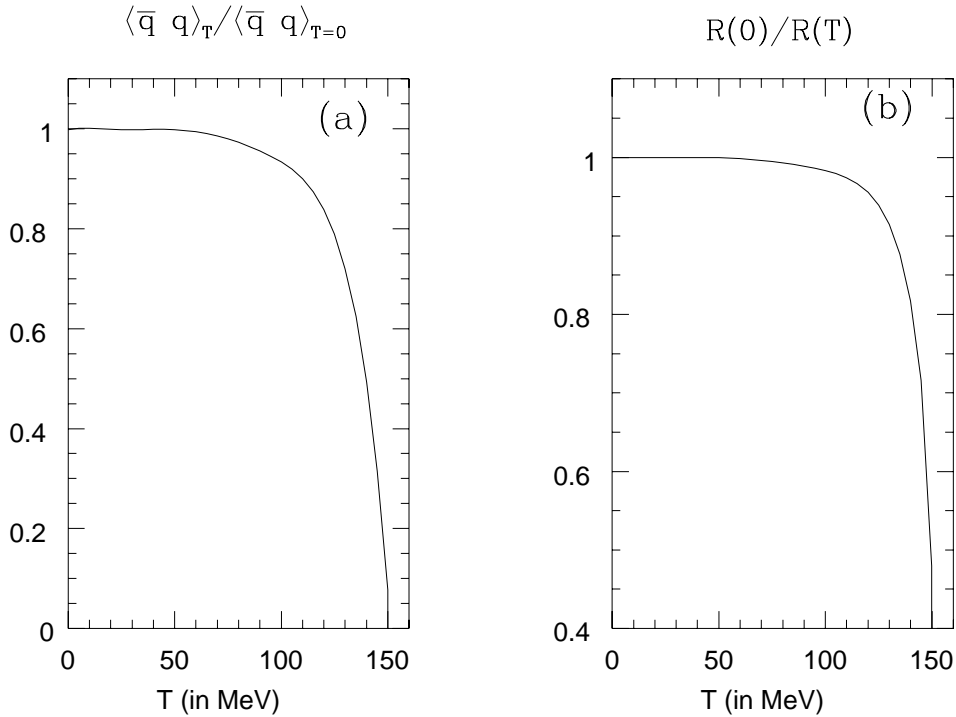}
     \caption
{\em {Figure (a) shows quark condensate at finite 
temperature normalised to that at zero temperature obtained from CHPT 
and Lattice. Figure (b) shows R(0)/R(T) as determined from Fig (a).}}
\end{figure}
       \label{condfig}
For intermediate regime we shall take a 
smooth interpolation between the two. The resulting behaviour of the 
temperature dependance of the  quark condensate and the ratio
$R(0)/R(T)$ determined using Eq.(\ref{qqrat}) are shown in
Fig.s \ref{condfig}.

With the temperature dependence of $R(T)$ known as above, the propagator
function of eq.(\ref{propt}) now completely defined. The finite temperature 
correlation function is now given as, parallel to Eq.(\ref{fcor})
\be
R(\zbf x,T)=-\left[Tr[S(\zbf x,T)\Gamma'S(-\zbf x,T)\Gamma]
+Tr\langle\left[\Sigma(\zbf x))\Gamma'\Sigma(-\zbf x)\Gamma
\right]\right]\rangle_T.
\label{fcort}
\ee
The second term in the above corresponds to contributions from
the fluctuations at finite temperature that can be written interms
of the functions $g_S(\zbf x,T), g_V(\zbf x, T)$ as earlier but now
being temperature dependent. 
We do not know how to calculate it except
for a general property that the effect of the four point structure
should decrease with temperature. We take here a simple ansatz for 
the temperature dependence of $g_V$ and $g_S$,
\be
g_{S,V}(\zbf x,T)=\left(\frac{\langle \bar q q\rangle_T}{\langle \bar q
q\rangle_{T=0}}\right)^2 \; \; g_{S,V}(\zbf x,T=0)
\label{gsvt}
\ee
The parameters $\mu_i$ are choosen to have the same values as of zero temperature while fitting the mesonic and baryonic correlation function. The normalised
correlation functions $R(\zbf x,T)/R_0(\zbf x,T=0)$ are plotted in Fig.4a
and Fig 4b respectively. 

As expected (on physical grounds) the amplitude of the correlator
decreases with increasing temperature. The peak of the vector
correlator shifts towards the right after $T=0.9T_c$. We might
remind ourselves that the position of the peak of the correlator is
inversely proportional to the mass of the particle in the relevant
channel\cite{neglecor}.

To extract hadronic properties at finite temperature, the correlators
are parametrised in terms of a spectral density function. 
This is a generalisation of eq.(\ref{rhodis}) to finite temperature 
and is given as
\cite{hatsuda,hatsuda2},
\be
\rho^V(s) = 3 \lambda_\rho^2 \delta (s-M_{\rho}^2) + 
\frac{3s}{4 \pi^2}\tanh \left[\frac{\sqrt{s}}{4T}\right]\theta (s-s_o)  
+ T^2 {S_\rho}\delta (s)
\label{specvec}
\ee
\be
\rho^P(s) = \lambda_\pi^2 \delta (s-M_\pi^2) + \frac{3s}{8
\pi^2} \tanh \left[\frac{\sqrt{s}}{4T}\right]
\theta (s-s_o) 
\label{specps}
\ee
\begin{figure}
\epsfbox[22 360 531 653]{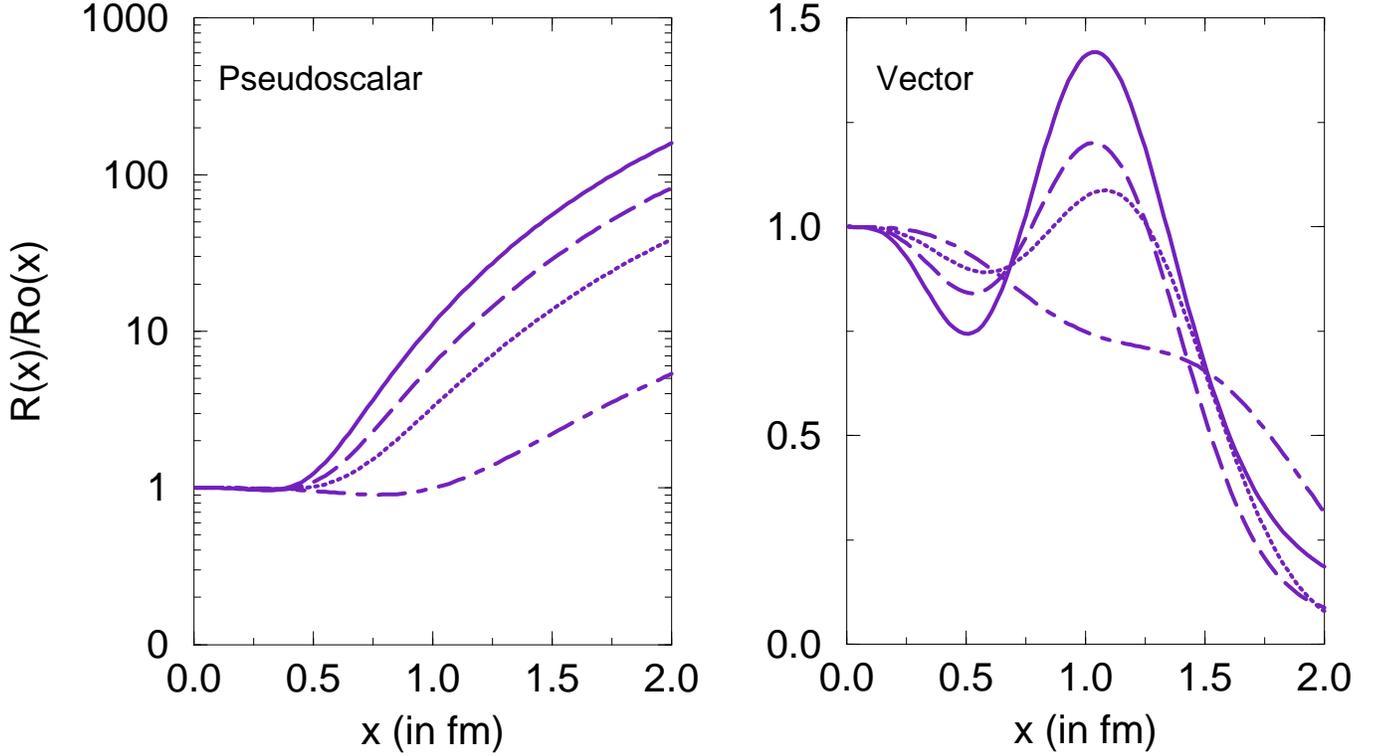}
\caption{\em The ratio of the meson correlation functions 
at finite
temperature to the correlation functions for noninteracting massless
quarks at zero temperature
$\em {R(x,T)}/{R_{0}(x,T=0)}  $, vs. distance x (in fm). 
The solid, dashed, dotted and dot-dashed lines correspond to temperatures 
$T=$0 MeV, $T=$130 MeV, $T=$140 MeV and $T=$148 MeV respectively.}
\label{corall}
\end{figure}

Eq.(\ref{specps}) corresponds to  spectral density function for
pseudoscalar channel. The last term
in Eq.(\ref{specvec})
The last term in  Eq.~(\ref{specvec})
is the scattering term for soft thermal
dissociations (mainly through pions), which exists only at finite
temperature\cite{hatsuda} and can be taken as
~\cite{hatsuda}$S_\rho\approx \frac{T^2}{9}$.

The mass, threshold and coupling are then extracted 
and the results are plotted
in Fig.~\ref{vecpar} for the vector and in 
the pseudoscalar
channel.
As can be seen from Fig.s\ref{corall}, with increase in
temperature, the correlation functions have a lower peak indicating
lack of correlations  with temperature. In the vector channel
the mass of the $\rho$ meson appears to decrease  for temperatures
beyond 120 MeV. The 
threshold for the continuum also decreases around the same temperature.
The behaviour with temperature of these quantities is qualitatively similar
to that found by Hatsuda {\it et al}\cite{hatsuda2}.
We have also plotted the temperature dependence of the coupling of the
bound state to the current which decreases with temperature but rather slowly
as compared to mass or the threshold for the
continuum. The temperature dependence of these parameters can be
used to calculate the lepton pair production rate from $\rho$ in the
context of ultra relativistic heavy ion collision experiments to
estimate vector meson mass shift in the medium. 

In the pseudoscalar channel the mass remains almost constant till the
critical temperature whereas the thershold and the coupling decrase with
the temparature. We have found  that in the 
pseudoscalar channel, the contribution to the correlation function
mostly comes from the fluctuating fields. Further, the temperature behaviour 
as taken in Eq.(\ref{gsvt})  essentially does not shift the position of 
the peak whereas the magnitude of the correlator decreases. That is 
reflected in the above behaviour of the parameters in 
the pseudoscalar channel. We may note here that similar behaviour of
pion mass becoming almost insensitive to temperature below the critical 
temperature was also observed in Ref.\cite{hatptp} where correlation functions
were calculated in a QCD motivated effective theory namely the 
Nambu- Jona Lasinio model.
\begin{figure}[h]
\epsfysize=10cm
\epsfbox[45 30 525 700]{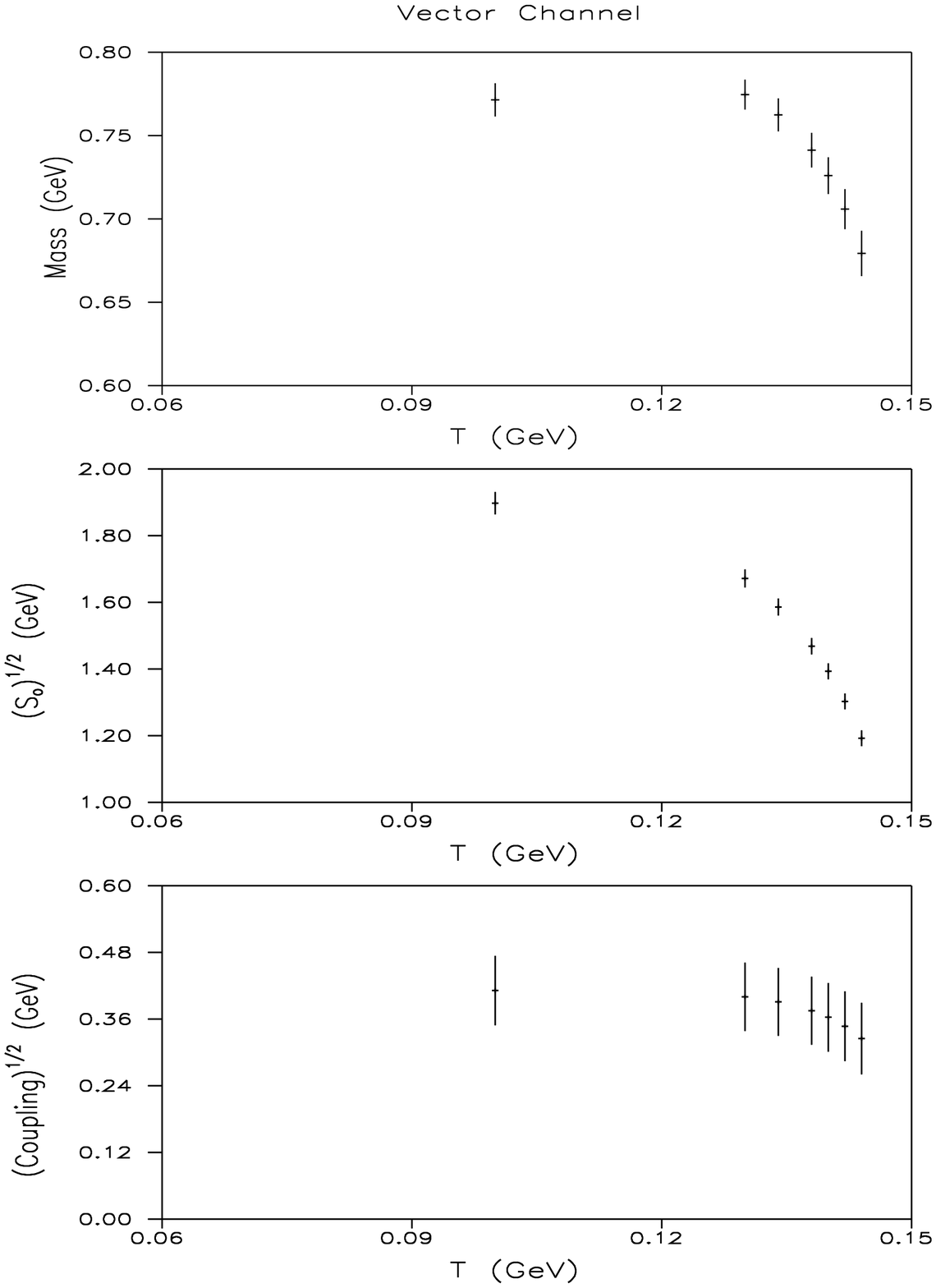},
\epsfysize=10cm
\epsfbox[53 30 525 700]{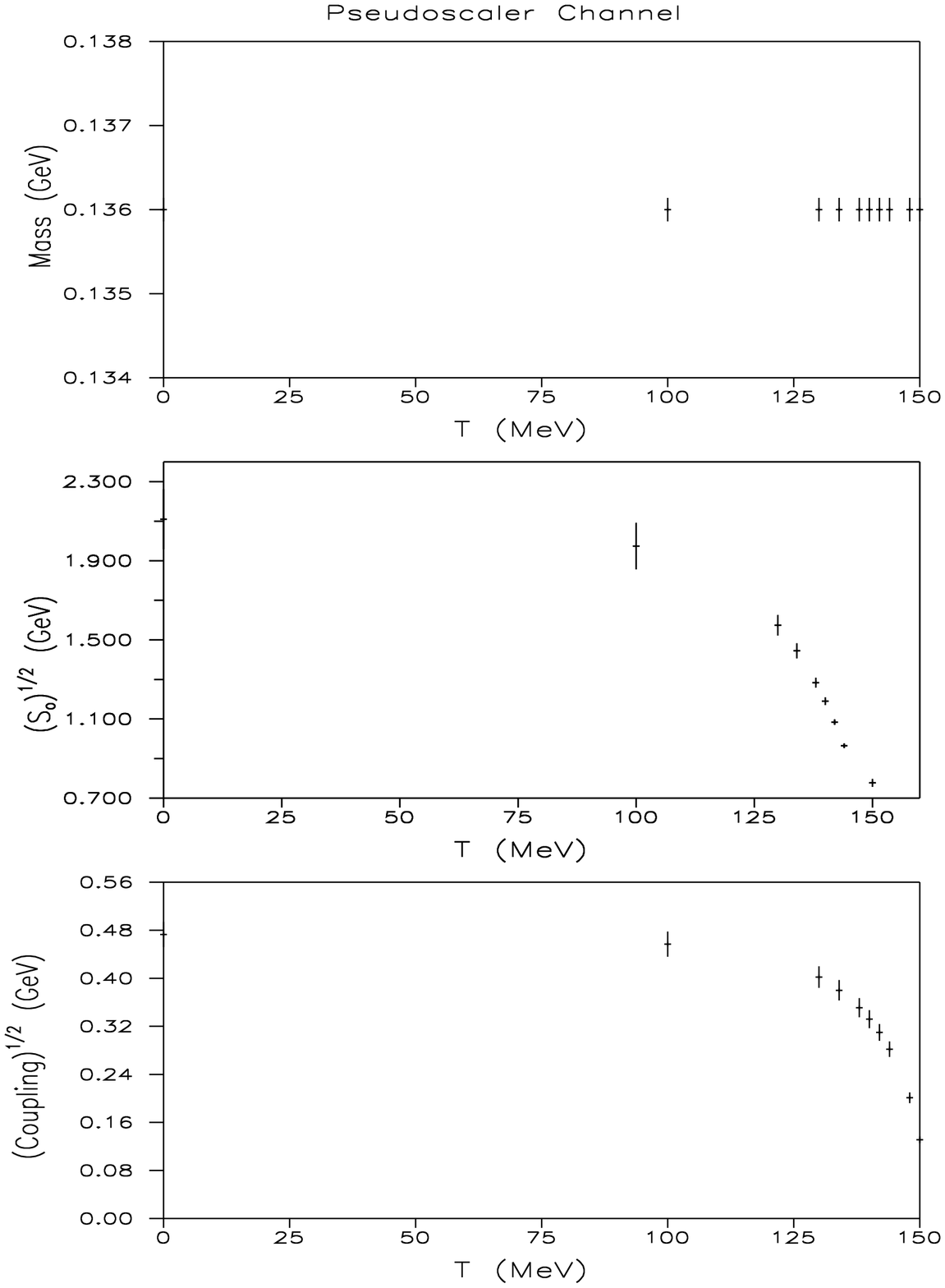}
\caption{\em The temperature dependence of mass, threshold (S$_0$) and
coupling for the vector (left panel) and pseudoscalar (right panel)
channel. T$_c=$150 MeV.
The vertical lines represent the errors obtained while fitting.}
\label{vecpar}
\end{figure}

\section{QCD vacuum at finite densities}
Let us now go over to explore the ground state structure in high
density QCD. Unlike high temperatures, rather little is known about
QCD at finite baryon densities from first principles like lattice
QCD simulation due to technical problems. However, different low energy
models of QCD seem to indicate a rich phase structure in this domain.
In particular the color superconductivity phase at high density
has attracted much attention. Although it was known for quite sometime
\cite{balin}, recent studies \cite{wil,sursc,pisarski,dia} indicated that
the superconducting gap could be as large as 100 MeV and  that has lead to
an extensive literature on this subject in recent past regarding its
 consequences in neutron stars as well as for heavy ion collisions.

As per BCS theorem, an arbitrarily weak attractive interaction makes the 
fermi sea of quarks unsatble at high densities. In deed, the color current
current interaction is attractive in the scalar $(\bar{\bfm 3})$ and
axial vector channel. This can lead to formation of Cooper pairs and
associated superconductivity in the color space. To discuss 
it in a nonperturbative manner, we take the trial ansatz over the 
chiral condensed vacuum
of Eq.(\ref{vac}) as \cite{jcphm}
\be
|\omega\rangle
=\exp\left[\frac{1}{2}\int \left[q_r^{ia}(\zbf k)^\dagger
f(\zbf k) q_{-r}^{jb}(-\zbf k)^\dagger
\epsilon_{ij}\epsilon_{3ab}
+\tilde q_r^{ia}(\zbf k)
f_1(\zbf k) \tilde q_{-r}^{jb}(-\zbf k)
\epsilon_{ij}\epsilon_{3ab}\right]
d\zbf k.\right]|vac\rangle
\label{vecd}
\end{equation}
\noindent 
In the above, $i,j$ are flavor indices, $a,b$ are the
color indices and $r(=\pm 1/2) $ is the spin index. We 
shall consider here two flavor and SU(3) color. 
 Here we have also introduced two trial functions $f(\zbf{k})$ and
$f_1(\zbf k)$ respectively for the diquark and diantiquark
channel. As may be noted the state constructed in Eq.(\ref{vecd}) is spin
singlet and is antisymmetric in color and flavor. 

Next, to include the effect of temperature and density, we obtain the 
state at finite temperature and density $|\omega(\beta,\mu)\rangle$ by a thermal
Bogoliubov transformation over the state $|\omega\rangle$ using thermofield
dynamics as  in Eq.(\ref{tvac})\cite{cort,umezawa,amph4,qcdtb},
\begin{equation} 
|\omega(\beta,\mu)\rangle=
\exp\left(
\int q_I (\zbf k)^\dagger \theta_-(\zbf k, \beta,\mu)
\underline q_I(\zbf k)^\dagger +
\tilde q_I (\zbf k) \theta_+(\zbf k, \beta,\mu)
\underline { \tilde q}_I (\zbf k) d\zbf k - h.c.\right)|\omega\rangle
\label{bth}
\end{equation}
In the above, 
the ansatz functions $\theta_{\pm}(\zbf k,\beta,\mu)$, as before,
will be related to quark and antiquark distributions.
We might note here that the trial ansatz given 
in Eq.(\ref{bth}) actually invloves five functions - $h(\zbf k)$,
for the quark anti quark condensates, $f(\zbf k)$ and $f_1(\zbf k)$ describing
respectively the diquark and diantiquark condensates and 
$\theta_{\pm}(\zbf k,\beta,\mu)$ to include the temperature and 
density effects. All these functions are to be obtained by minimising the
thermodynamic potential. This will involve an assumption about the effective 
hamiltonian . For the purpose of illustration we shall consider a
hamiltonian of Nambu-Jonalasinio type given as

\be
{\cal H}=\psi^\dagger(-i\bfm \alpha \cdot \bfm \nabla )\psi
+\frac{g^2}{2}J_\mu^aJ^{\mu a}.
\label{ham}
\ee

with $J_\mu^a =\bar q \gamma^\mu T^a q$. One can then calculate the
energy functional
$\epsilon=\langle \omega(\beta,\mu)|{\cal H}|\omega(\beta,\mu)\rangle$ and
the thermodynamic potential ${\cal F}=\epsilon-(1/\beta)S -\mu N$, using the
Bogoliubov technique. The details are given in Ref. \cite{jcphm}. The
free energy is a functional of all the five functions $h,f,f_1,\theta_{\pm}$.
However with the point interaction of the the Hamiltonian of eq.(\ref{ham})
one can {\em determine} them. This results in two coupled gap equations
to be solved in a self consistent manner and are given as
\begin{mathletters}

\be
\frac{4g^2}{3}\frac{1}{(2\pi)^3}
\int d\zbf k \frac{1}{\sqrt{\zbf k^2+M^2}}
\left(\frac{\xi_-}{\omega_-}tan h(\frac{\beta\omega_-}{2}) 
+\frac{\xi_+}{\omega_+}tan h(\frac{\beta\omega_+}{2})
\right)=1
\label{mgap1}
\ee 
\be
\frac{4g^2}{3}\frac{1}{(2\pi)^3}
\int d\zbf k \left(\frac{tan h(\frac{\beta\omega_-}{2})}{\omega_-}
+ \frac{tan h(\frac{\beta\omega_+}{2})}{\omega_+}
\right)=1
\label{scgap1}
\ee 
\end{mathletters}
where, $\omega_{\pm}=\sqrt{\Delta^2+\xi_{\pm}^2}$ and 
$\xi_{\pm}=(E\pm\nu)$. Here, $E$ is the energy of the quasi paricles
given as $E=\sqrt{(k^2+M^2)}$, and,$\nu$ is the chemical potential in
presence of interaction given as $\nu=\mu-(4g^2/3)(N/12)$, $N$ being the
quark number density. Eq.(\ref{mgap1}) is the mass gap equation
in presence of diquark condensates and Eq.(\ref{scgap1}) is the
superconducting gap equation which is a relativistic
generalisation of BCS gap equation \cite{bcs,sarah}. 
In the limit of no diquark condensates Eq.(\ref{mgap1})
reduces to that obtained in Ref.\cite{sarah} except for the
numerical factors before the integrand. This is due to the fact that 
the approximation the the later case has been a mean field 
approximation \cite{asakawa} unlike the case here where
 approximation lies
only with the ansatz for the ground state. Eq.(\ref{mgap1})
without the diquark condensate contributions
is also have the same structure as in Ref.\cite{wil} in the limit of the
formfactors intoduced in the later case reduces to a 
sharp cutoff in the momentum. Similarly in the limit of chiral
condensate going to zero, the super conducting gap equation 
(\ref{scgap1}) is similar to that obtained in Ref.\cite{sarah}
or Ref.\cite{wil}.
\begin{figure}
\begin{center}
\epsfig{file=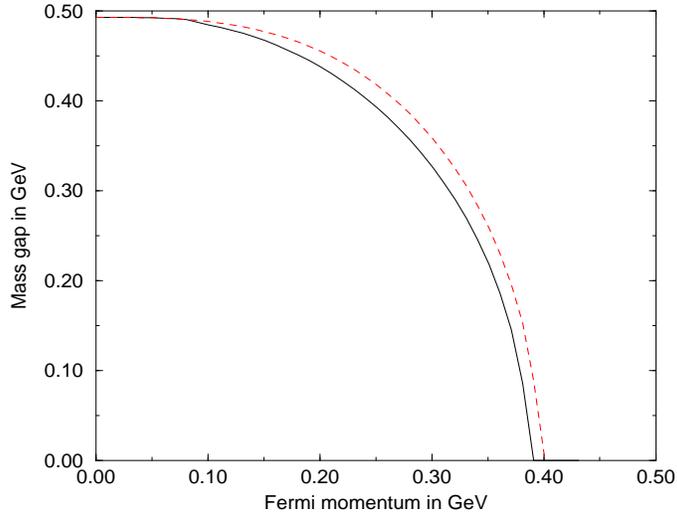,width=10cm,height=8cm}
\end{center}
\caption{\em Mass gap $M$ as a function of fermi mometum. The dashed line
correspond to no diqurak condensates. The solid line corresponds to
both diquark and quark antiquark condensates structure for the ground state.}
\label{dmmgapfig}
\end{figure}

The solution of these equation
at zero temperature is shown in Fig.(\ref{dmmgapfig}) for the mass
gap and in Fig.(\ref{dmsugapfig}) for the superconducting
gap. For the sake of comparision we have also plotted the mass
gap without the diquark condensates. The coupling here taken as
$g^2\sim 56 GeV^{-2}$ and the cutoff$\Lambda\sim 0.67 GeV$. these
values are taken so as to give the same transition temperature
as in ref\cite{wil,sarah}. With these couplings, the mass gap at zero 
temperature and density is about 490 MeV and the mass gap vanishes 
at fermi momentum  $k_f=\sqrt{\nu^2-M^2} =400 MeV$. The corresponding
critical quark number density is about $1.7/fm^3$.
\begin{figure}
\begin{center}
\epsfig{file=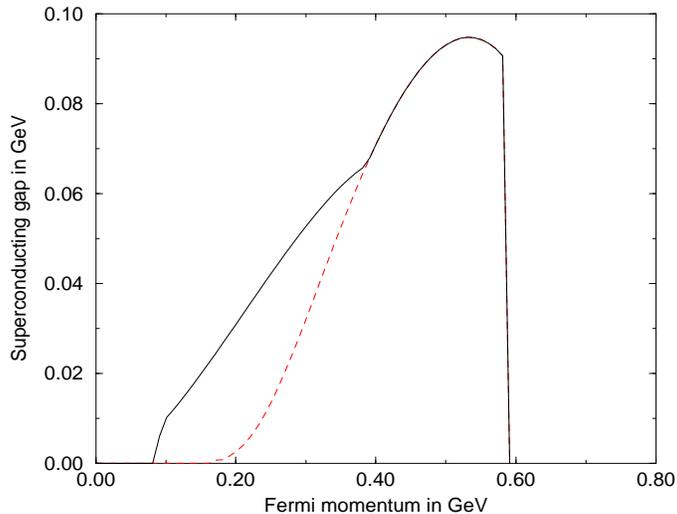,width=10 cm,height=8cm}
\end{center}
\caption{\em Superconducting gap $\Delta$
as a function of fermi mometum. The solid line corresponds to 
both diquark and quark antiquark condensates. 
The dashed line corresponds to only diquark condensates. }
\label{dmsugapfig}
\end{figure}

 Presence of 
diquark condensate does not change these values very much. 
The diquark condensate increases with number density and 
becomes maximum of about 90 MeV beyond which the 
effect of the cutoff is felt and it vanishes 
for fermi momentum  around 600 MeV. We also plot the equation of state (EOS)
in Fig.\ref{dmprefig}.
\begin{figure}
\begin{center}
\epsfig{file=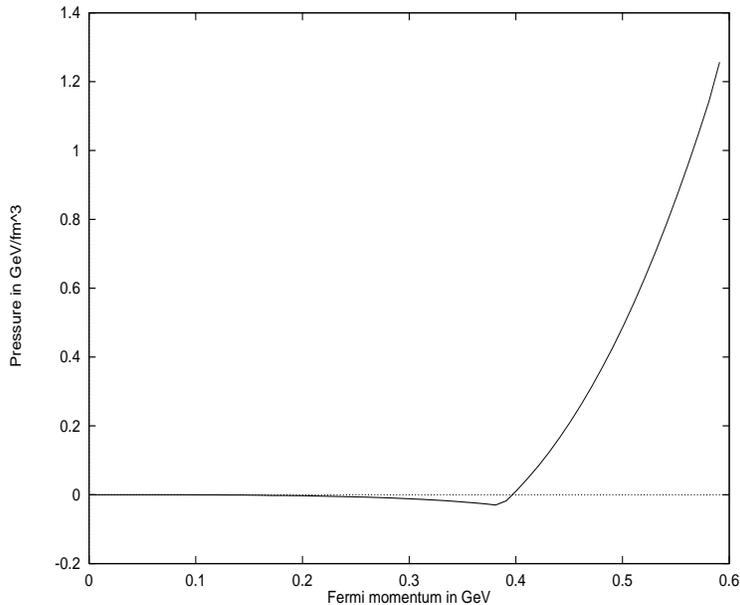,width=10 cm,height=8cm}
\end{center}
\caption{\em Pressure as a function of fermi momentum . }
\label{dmprefig}
\end{figure} 
While plotting the EOS, we have added the bag constant $\epsilon_0$,
which is the difference in energy density of the perturbative vacuum
and the nonperturbative vacuum with condensate at zero temperature
 and density, to the pressure. With the present parameters the bag constant 
 turns out to be 
$-(200MeV)^4$. The pressure has a cusp like structure and becomes negative
at finite densities. The portion of the curve that goes down with $k_f$
corresponds to the nontrivial solution to the mass gap and the portion 
that increases with density correspond to zero mass solution. The
negative pressure indicates mechanical instability and can have the 
interpretation that uniform nonzero density matter will break up in to droplets
of finite density in which chiral symmetry is restored surrounded
by empty space with zero pressure and density \cite{wil}. 
It is tempting to identify
the droplets of quark matter with nucleons within which density is nozero and
$\langle \bar q q\rangle=0 $ -- a  fact reminscent of bag models \cite{wil}. 
Nothing in the model however says that the droplets have quark number three.
The equation of state does not change
much in presence of diquark condensates. This should be expected
because the effect of diquark condensate is small in the region where
chiral condensate is non vanishing. Thus gross structural properties of the 
neutron stars are not likely to be affected by diquark condensates. However,
cooling of neutron stars  shall be expected to 
be very much affected by such a gap of about 100 MeV.
\section{summary and discussions}

We have looked into the structure of QCD vacuum in a nonperturbative
manner with a variational ansatz. This has been done for zero temperature
as well as finite temperature and densities. The input has been equal time
algebra for the field operators and the ansatz for the ground state.
At zero temperature, the Bogoliubov type pairing ansatz involving
{\em both} quark and gluon condensates becomes energetically favourable
beyond a critical coupling. It also gave some of the low energy
hadronic properties like pion and proton charge raddi, proton,neutron magnetic
moments and the bag constant to be around their phenomenological values
for  the coupling value of $1.28$. Thus $\alpha_s=1.28$
 effectively corresponds to the QCD coupling constant for the vacuum
 configuration. With optimized renormalisation group equations, it has been
seen that \cite{stevenson} $\alpha_s(Q)$ {\em does not } go to infinity as $Q$
decreases below 300 MeV, but freezes to a constant value around
unity. Our analysis seems to remind us of a similar situation.

 We next evaluated the correlation functions
of hadronic currents in such a condensed medium. However it appears that
to have quantitative agreement with correlator phenomenology, particularly
in the pseudoscalar channel, these cannot be described by the propagators
alone but must of necessity have the fluctuations of the condensates.
This may be looked upon as combination
of two effects --(i) an effective way of incorporating gluon condensate effects
and (ii)the existance of explicit four point structure in QCD vacuum.
In some ways these fluctuations may be related to the ``hidden contributions"
discussed by Suryak \cite{suprop}.

It is worthwhile pointing out that OPE and our approach are based on
intrinsically different assumptions. The former is an expansion
which separates short distance (Wilson coefficients) and long distance
(condensates) physics. In  our method we assume an explicit vacuum structure
in terms of quark condensate (two point function) plus an irreducible four
point function. Having made an ansatz for the vacuum,
we do not make any further approximation in the evaluation of the correlators.
The approach is phenomenological in the sense that the values of the parameters in the four point function \cite{plb} are chosen to reproduce the behaviour 
of the correlators. 
As emphasized by Shuryak \cite{surcor} and Sch$\ddot{a}$fer and Shuryak
\cite{surcor2}
OPE is able to quantitatively describe the zero temperature pion correlation
functions for small x (upto 0.25 fm)  but underpredicts it for large x. 
In our work\cite{propcor,plb} the agreement is quantitative with 
experimentally deduced mesonic correlation function and
lattice results \cite{neglecor} for the whole range. 
This covers the small x values, resonance region and large x domain where
 the correlator vanishes.  We have however more parameters. To the extent
that the large x behaviour of the pionic correlator depends on gluon
condensates in OPE our parametrization would imply an effective way 
of including gluon condensate effects.

 For studying the correlators at finite 
temperature, we assume that the two point as well as the four point 
function vanish at $T=T_C$. The parameters in the four point function 
are assumed constant at their $T=0$ values -- no additional T dependence
is given to them.
We see that our results are broadly in agreement with those of Hatsuda etal
\cite{hatsuda2}. All the same, it is not possible to carry out a term by term 
comparison of our results with OPE results. This is because the assumptions are
different in the two approaches.
We recall that Hatsuda etal \cite{hatsuda2}
attribute the decrease in the $\rho$ mass to contributions
coming from four point function. They also find variation in the gluon
condensates  to be less than 5{\%} over the temperature region
that is considered.
In our work  we do not include the gluon condensates and
the four point structure vanishes as $T\rightarrow T_C$. This is
not the case for gluon condensate ($\langle GG\rangle$) in lattice 
calculations or in the dilute pion gas model of Ref. \cite{hatsuda2}.
Consequently, we may be tempted to infer that 
the decrease in rho mass in our model is due to ``genuine" four point
function effects (as in Hatsuda etal \cite{hatsuda2})
and not from the ``effective" gluon condensate contribution. One,
however, has to be very cautious. This is because in the present work
 the parameters in the
four point function were kept constant at their $T=0$ values. It should be
recalled that these values were determined so as to correctly 
describe the behaviour of the correlators  (in particular pion) at T=0.
Hence the parameters do reflect some effective gluon condensate effects. 
The results at finite temperature would certainly be modified if
these parameters are given significant T dependence. Our parametrization 
is such that if the parameters decrease with T then the contribution
to the correlator will decrease.
The crucial question however still remains as to the behaviour of 
$\langle GG\rangle$ for $T<T_C$. If it varies very little in this range 
then our assumption of the parameters remaining constant would be
reasonable.

We would like to add here that the present analysis will be valid 
for temperatures below the critical temperature. Above the critical
temperature there have been calculations essentially using finite
temperature perturbative QCD in random phase approximations (RPA)
\cite{parikh}. 
However, in the region above $T_C$,
nonperurbative features have been known to exist from studies
in lattice QCD simulations\cite{lman}. 
In view of this, one may have to carry out a 
hard thermal loop calculation where a partial resummation is 
done\cite{brat}. Alternatively, one may use other nonperturbative
approaches such as QCD sum rules at finite temperature \cite{hanson} or RPA
approach in an instanton liquid model for the QCD vacuum \cite{vel}.

The vacuum structure at finite densities was looked into
taking into account the possibility of diquark condensates
in a Nambu-Jonalasinio model. As earlier, the approximation
lies here only in the ansatz. The resulting gap equation for
the superconducting gap is a relativistic generalisation
of the BCS gap equation. The mass gap equation in the presence of
superconducting gap is a new feature of the present calculation.
Because of the point interaction as in ${\cal H}_{int}$ of
Eq.(\ref{ham})we could solve for the gap functions explicitly.
In the presence of realistic potentials as in Ref.\cite{alkz}or
Ref.\cite{qcdtb} one will have to solve an integral equation for
the gap functions. Such a calculation is in progress \cite{jcphm}.
It will be interesting to see how the results for
the superconducting gap would be affected in presence of 
quark antiquark condensates as compared to the resummed perturbative QCD
calculations at finite densities \cite{pisarski}. The equation of state
here did not change very much. Thus the global properties of neutron stars
shall not be affected in an appreciable manner. However, a gap of about
 100 MeV can have its implications on neutron star cooloing \cite{blashe},
magnetic fields of pulsars and their thermal evolution \cite{sedra}.
Future theoretical studies of QCD at finite baryon densities may reveal
to which extent these newly established features of quark matter
considered in effective models shall have their correspondence in a more
complete treatment of QCD at finite baryonic density and shed light on whether
one may expect any distinguishing feature in the global properties
of neutron/quark stars.

\acknowledgements I would like to thank S.P. Misra,
Amruta Mishra, J.C. Parikh and Varun Sheel for an enjoyble and
fruitful collaboration over the  years.

\def \feyn{R.P. Feynman, Nucl. Phys. B 188, 479 (1981) }
\def\parikh {Jitendra C. Parikh, Philip J. Siemens, Phys. Rev. D37, 3246
(1988); R.B. Thayyullathil and J.C. Parikh, Phys. Rev. D44, 3964, (1991).}
\def\lman {E. Laermann, Nucl. Phys. A610, 1c (1996).}
\def\blashe{D. Blashke, T. Klaehn and D.N. Voskorensky,astro-ph9908334.}
\def\sedra{D. Blashke, D.M. Sedrakian and K.M. Shahabasyan, astro-ph9904395,
M.Alford, J. Berges and K.Rajagopal,astro-ph/9910254.}
\def\brat
{Eric Braaten and Robert D. Pisarski, Phys. rev. Lett. 64, 1338 (1990),
Nucl. Phys. B339, 310 (1990).}
\def\hanson
{J. Hansson and I. Zahed, {\it ``QCD sumrules at high 
temperature"}, SUNY preprint, NTG 90-338 (1990).}
\def\vel{M.Velkovsky and E.V. Shuryak, Phys. Rev. D56, 2766 (1996).}
\def\suryak{R.Rapp,T.Shaeffer,E.Shuryak and M. Velkovsky, Phys. Rev.
Lett. 81,53 (1998).}
\def\dia{D.I. Diakonov, H. Forkel, M.Lutz, Phys. Lett.B373,147 (1996).}
\def\balin{D.Ballin and A. Love, Phys. rep. 107,325 (1984).}

\def \svz {M.A. Shifman, A.I. Vainshtein and V.I. Zakharov,
Nucl. Phys. B147, 385, 448 and 519 (1979);
R.A. Bertlmann, Acta Physica Austriaca 53, 305 (1981)}

\def \spmbst {S.P. Misra, Phys. Rev. D35, 2607 (1987)}

\def \hmgrnv { H. Mishra, S.P. Misra and A. Mishra,
Int. J. Mod. Phys. A3, 2331 (1988)}
\def\umezawa
 {H.~Umezawa, H.~Matsumoto and M.~Tachiki {\it Thermofield dynamics
and condensed states} (North Holland, Amsterdam, 1982) ;
P.A.~Henning, Phys.~Rep.253, 235 (1995).}
\def\hatptp{T. Hatsuda and T. Kunihiro, 
Prog. Theor. Phys. Suppl. 91, 284 (1987).}

\def\jcphm {H.Mishra and J.C. Parikh - in preparation}
\def \snss {A. Mishra, H. Mishra, S.P. Misra
and S.N. Nayak, Phys. Lett 251B, 541 (1990)}

\def \amqcd { A. Mishra, H. Mishra, S.P. Misra and S.N. Nayak,
Pramana (J. of Phys.) 37, 59 (1991). }
\def\qcdtb{A. Mishra, H. Mishra, S.P. Misra 
and S.N. Nayak, Z.  Phys. C 57, 233 (1993); A. Mishra, H. Mishra
and S.P. Misra, Z. Phys. C 58, 405 (1993)}

\def \spmtlk {S.P. Misra, Talk on {\it `Phase transitions in quantum field
theory'} in the Symposium on Statistical Mechanics and Quantum field theory, 
Calcutta, January, 1992, hep-ph/9212287}

\def \hmnj {H. Mishra and S.P. Misra, Phys. Rev. D 48, 5376 (1993)}

\def \hmqcd {A. Mishra, H. Mishra, V. Sheel, S.P. Misra and P.K. Panda,
hep-ph/9404255 (1994)}

\def \amcrl {A. Mishra, H. Mishra and S.P. Misra, Z. Phys. C 57, 241 (1993)}

\def \higgs { S.P. Misra, in {\it Phenomenology in Standard Model and Beyond}, 
Proceedings of the Workshop on High Energy Physics Phenomenology, Bombay,
edited by D.P. Roy and P. Roy (World Scientific, Singapore, 1989), p.346;
A. Mishra, H. Mishra, S.P. Misra and S.N. Nayak, Phys. Rev. D44, 110 (1991)}

\def \nmtr {A. Mishra, 
H. Mishra and S.P. Misra, Int. J. Mod. Phys. A5, 3391 (1990); H. Mishra,
 S.P. Misra, P.K. Panda and B.K. Parida, Int. J. Mod. Phys. E 1, 405, (1992);
 {\it ibid}, E 2, 547 (1993); A. Mishra, P.K. Panda, S. Schrum, J. Reinhardt
and W. Greiner, to appear in Phys. Rev. C}

\def \dtrn {P.K. Panda, R. Sahu and S.P. Misra, 
Phys. Rev C45, 2079 (1992)}

\def \qcd {G. K. Savvidy, Phys. Lett. 71B, 133 (1977);
S. G. Matinyan and G. K. Savvidy, Nucl. Phys. B134, 539 (1978); N. K. Nielsen
and P. Olesen, Nucl.  Phys. B144, 376 (1978); T. H. Hansson, K. Johnson,
C. Peterson Phys. Rev. D26, 2069 (1982)}

\def \cornwal {J.M. Cornwall, Phys. Rev. D26, 1453 (1982)}

\def \mndglv {J. E. Mandula and M. Ogilvie, Phys. Lett. 185B, 127 (1987)}

\def \schwinger {J. Schwinger, Phys. Rev. 125, 1043 (1962); ibid,
127, 324 (1962)}

\def \schutte {D. Schutte, Phys. Rev. D31, 810 (1985)}

\def \amspm {A. Mishra and S.P. Misra, Z. Phys. C 58, 325 (1993)}

\def \gft{ For gauge fields in general, see e.g. E.S. Abers and 
B.W. Lee, Phys. Rep. 9C, 1 (1973)}

\def \gribov {V.N. Gribov, Nucl. Phys. B139, 1 (1978)}

\def \spm78 {S.P. Misra, Phys. Rev. D18, 1661 (1978); {\it ibid}
D18, 1673 (1978)} 

\def \lopr {A. Le Youanc, L.  Oliver, S. Ono, O. Pene and J.C. Raynal, 
Phys. Rev. Lett. 54, 506 (1985)}

\def \spphi {S.P. Misra and S. Panda, Pramana (J. Phys.) 27, 523 (1986);
S.P. Misra, {\it Proceedings of the Second Asia-Pacific Physics Conference},
edited by S. Chandrasekhar (World Scientfic, 1987) p. 369}

\def\spmdif {S.P. Misra and L. Maharana, Phys. Rev. D18, 4103 (1978); 
    S.P. Misra, A.R. Panda and B.K. Parida, Phys. Rev. Lett. 45, 322 (1980);
    S.P. Misra, A.R. Panda and B.K. Parida, Phys. Rev. D22, 1574 (1980)}

\def \spmvdm {S.P. Misra and L. Maharana, Phys. Rev. D18, 4018 (1978);
     S.P. Misra, L. Maharana and A.R. Panda, Phys. Rev. D22, 2744 (1980);
     L. Maharana,  S.P. Misra and A.R. Panda, Phys. Rev. D26, 1175 (1982)}

\def\spmthr {K. Biswal and S.P. Misra, Phys. Rev. D26, 3020 (1982);
               S.P. Misra, Phys. Rev. D28, 1169 (1983)}

\def \spmstr { S.P. Misra, Phys. Rev. D21, 1231 (1980)} 

\def \spmjet {S.P. Misra, A.R. Panda and B.K. Parida, Phys. Rev Lett. 
45, 322 (1980); S.P. Misra and A.R. Panda, Phys. Rev. D21, 3094 (1980);
  S.P. Misra, A.R. Panda and B.K. Parida, Phys. Rev. D23, 742 (1981);
  {\it ibid} D25, 2925 (1982)}

\def \arpftm {L. Maharana, A. Nath and A.R. Panda, Mod. Phys. Lett. 7, 
2275 (1992)}

\def \van {R. Van Royen and V.F. Weisskopf, Nuov. Cim. 51A, 617 (1965)}

\def \rchpi {S.R. Amendolia {\it et al}, Nucl. Phys. B277, 168 (1986)}

\def \chrl{ Y. Nambu, Phys. Rev. Lett. \zbf 4, 380 (1960);
A. Amer, A. Le Yaouanc, L. Oliver, O. Pene and
J.C. Raynal, Phys. Rev. Lett.\zbf  50, 87 (1983);
ibid, Phys. Rev.\zbf  D28, 1530 (1983); 
M.G. Mitchard, A.C. Davis and A.J.
Macfarlane, Nucl. Phys. \zbf B325, 470 (1989);
B. Haeri and M.B. Haeri, Phys. Rev.\zbf  D43,
3732 (1991);; V. Bernard, Phys. Rev.\zbf  D34, 1601 (1986);
 S. Schram and
W. Greiner, Int. J. Mod. Phys. \zbf E1, 73 (1992)}
\def\finger{ J.R. Finger and J.E. Mandula, Nucl. Phys. \zbf B199, 168 (1982).}
\def\davis{S.L. Adler and A.C. Davis, Nucl. Phys.\zbf  B244, 469 (1984) .}
\def\alkofer{R. Alkofer and P. A. Amundsen, Nucl. Phys.\zbf B306, 305 (1988)}
\def\klev{S.P. Klevensky, Rev. Mod. Phys.\zbf  64, 649 (1992);}
 \def\bhalerao{S. Li, R.S. Bhalerao and R.K. Bhaduri, Int. J. Mod. Phys. 
 \zbf A6, 501 (1991)}
\def\asakawa{M. Asakawa and K. Yazaki, Nucl. Phys. A504,668 (1989).}

\def \spmijp { S.P. Misra, Ind. J. Phys. 61B, 287 (1987)}

\def \feynman {R.P. Feynman and A.R. Hibbs, {\it Quantum mechanics and
path integrals}, McGraw Hill, New York (1965)}

\def \glstn{ J. Goldstone, Nuov. Cim. \zbf 19, 154 (1961);
J. Goldstone, A. Salam and S. Weinberg, Phys. Rev. \zbf  127,
965 (1962)}

\def \anderson {P.W. Anderson, Phys. Rev. \zbf {110}, 827 (1958)}

\def \nambu{ Y. Nambu, Phys. Rev. Lett. \zbf 4, 380 (1960)}

\def\donogh {J.F. Donoghue, E. Golowich and B.R. Holstein, {\it Dynamics
of the Standard Model}, Cambridge University Press (1992)}

\def\satz {T. Matsui and H. Satz, Phys. Lett. B178, 416 (1986)}

\def\cps {C. P. Singh, Phys. Rep. 236, 149 (1993)}

\def\prliop {A. Mishra, H. Mishra, S.P. Misra, P.K. Panda and Varun
Sheel, Int. J. of Mod. Phys. E 5, 93 (1996)}

\def\hmcor {V. Sheel, H. Mishra and J.C. Parikh, Phys. Lett. B382, 173
(1996); {\it biid}, to appear in Int. J. of Mod. Phys. E}
\def\propcor
{Varun~Sheel, Hiranmaya~Mishra and Jitendra C.~Parikh,
Int. J. Mod. Phys. E6, 275, (1997).}
\def\suprop
{E.V.~Shuryak and J.J.M.~Verbaarschot, Nucl. Phys. B410, 37 (1993).}
\def\cort { V. Sheel, H. Mishra and J.C. Parikh, Phys. ReV D59,034501 (1999);
{\it ibid}Prog. Theor. Phys. Suppl.,129,137, (1997).}
\def\amph4{Amruta Mishra and Hiranmaya Mishra, J. Phys. G23,143, (1997).}

\def\surcor {E.V. Shuryak, Rev. Mod. Phys. 65, 1 (1993)} 

\def\stevenson {A.C. Mattingly and P.M. Stevenson, Phys. Rev. Lett. 69,
1320 (1992); Phys. Rev. D 49, 437 (1994)}
\def\plb{
Varun~Sheel, Hiranmaya~Mishra and Jitendra C.~Parikh,
Phys. Lett. B382, 173 (1996).}

\def\mac {M. G. Mitchard, A. C. Davis and A. J. Macfarlane,
 Nucl. Phys. B 325, 470 (1989)} 
\def\tfd
 {H.~Umezawa, H.~Matsumoto and M.~Tachiki {\it Thermofield dynamics
and condensed states} (North Holland, Amsterdam, 1982) ;
P.A.~Henning, Phys.~Rep.253, 235 (1995).}

\def \neglecor{M.-C. Chu, J. M. Grandy, S. Huang and 
J. W. Negele, Phys. Rev. D48, 3340 (1993);
ibid, Phys. Rev. D49, 6039 (1994)}

\def\revdata {Particle Data Group, Phys. Rev. D 50, 1173 (1994)}

\def\sinp {S.P. Misra, Indian J. Phys., {\bf 70A}, 355 (1996)}

\def\bryman {D.A. Bryman, P. Deppomier and C. Le Roy, Phys. Rep. 88,
151 (1982)}

\def\thooft {G. 't Hooft, Phys. Rev. D 14, 3432 (1976); D 18, 2199 (1978);
S. Klimt, M. Lutz, U. Vogl and W. Weise, Nucl. Phys. A 516, 429 (1990)}
\def\alkz { R. Alkofer, P. A. Amundsen and K. Langfeld, Z. Phys. C 42,
199(1989), A.C. Davis and A.M. Matheson, Nucl. Phys. B246, 203 (1984).}
\def\sarah {T.M. Schwartz, S.P. Klevansky, G. Papp, Phys. Rev. C60,055205
(1999).}
\def\wil{M. Alford, K.Rajagopal, F. Wilczek, Phys. Lett. B422,247(1998)i
{\it{ibid}}Nucl. Phys. B537,443 (1999).}
\def\sursc{R.Rapp, T.Schaefer, E. Suryak and M. Velkovsky Phys. Rev. Lett.
81, 53(1998),{\it {ibid}},hep-ph/9904353.}
\def\pisarski{D. Rische, R.D. Pisarski, Phys Rev D60,094013, 1999;{\it ibid}
nucl-th/9910056, nucl-th/9907041.}
\def\leblac {M. Le Bellac, {\it Thermal Field Theory}(Cambridge, Cambridge University
Press, 1996).}
\def\bcs{A.L. Fetter and J.D. Walecka, {\it Quantum Theory of Many
particle Systems}(McGraw-Hill, New York, 1971.}
\def\alexander{Aleksander Kocic, Phys. Rev. D33, 1785,(1986).}

\end{document}